\documentclass[11pt]{article}
\usepackage[margin=1in]{geometry}
\usepackage{amsmath,amssymb,amsthm, amsfonts}
\usepackage{etaremune, enumerate, float, verbatim}
\usepackage{setspace}
\usepackage{algorithm}
\usepackage{algorithmic}
\usepackage{hyperref}
\usepackage{graphicx}
\usepackage{setspace}
\usepackage{url}
\usepackage{float}
\usepackage{multirow}
\usepackage{spreadtab}
\usepackage{numprint}

\usepackage{color, colortbl}
\usepackage{subcaption, caption}
\usepackage[algo2e,ruled,vlined]{algorithm2e}
\usepackage{mathrsfs,amsmath}
\usepackage{color}
\usepackage{times}
\usepackage[colorinlistoftodos]{todonotes}

\title{Modeling of Network Based Digital Contact Tracing and Testing Strategies for the COVID-19 Pandemic}
\author{Daniel Xu}
\date{November 2020}

\begin{document}
\maketitle

\begin{abstract}

With more than 1.7 million COVID-19 deaths, identifying effective measures to prevent COVID-19 is a top priority. We developed a mathematical model to simulate the COVID-19 pandemic with digital contact tracing and testing strategies. The model uses a real-world social network generated from a  high-resolution contact data set of 180 students. This model incorporates infectivity variations, test sensitivities, incubation period, and asymptomatic cases. We present a method to extend the weighted temporal social network and present simulations on a network of 5000 students. The purpose of this work is to investigate optimal quarantine rules and testing strategies with digital contact tracing. The results show that the traditional strategy of quarantining direct contacts reduces infections by less than 20\% without sufficient testing. Periodic testing every 2 weeks without contact tracing reduces infections by less than 3\%. A variety of strategies are discussed including testing second and third degree contacts and the pre-exposure notification system, which acts as a social radar warning users how far they are from COVID-19. The most effective strategy discussed in this work was combined the pre-exposure notification system with testing second and third degree contacts. This strategy reduces infections by 18.3\% when 30\% of the population uses the app, 45.2\% when 50\% of the population uses the app, 72.1\% when 70\% of the population uses the app, and 86.8\% when 95\% of the population uses the app. When simulating the model on an extended network of 5000 students, the results are similar with the contact tracing app reducing infections by up to 79\%.
\end{abstract}

\section{Introduction}\label{intro}
More than 1.7 million people have died of COVID-19 \cite{covidmap}. With such catastrophic loss of life at risk, it is a top priority to identify effective measures to prevent the spread of COVID-19. Preventing the spread of COVID-19 has several challenges. Firstly, 44\% of COVID-19 transmissions occur during the asymptomatic stage with infected individuals being most infectious 1 to 2 days before symptom onset \cite{ViralShedding}. Secondly, according to the CDC Planning Scenarios report \cite{CDCScenarios}, 40\% of all COVID-19 cases remain asymptomatic and this number could be as high as 79\% for those under 20 \cite{ChildrenAsymp}. The CDC Planning Scenario also estimates that asymptomatic individuals are 75\% as infectious as symptomatic individuals. Finally, the false-negative rate of RT-PCR tests on presymptomatic individuals ranges from 67\% to 100\% \cite{testSensitivity}. All of these factors make it very difficult to prevent COVID-19 with symptom-based measures.

Contact tracing has existed for decades, helping to reduce tuberculosis \cite{Tuberculosis}, sexually transmitted diseases \cite{STD}, and Ebola \cite{Ebola}. However, manual contact tracing relies on substantial human labor. According to a recent survey by NPR, 39 states do not have enough contact tracers \cite{NPRSurvey}. Additionally, of the contacts reported, only around 50\% were reachable by contact tracers \cite{ManualCTHard}. Health officials estimate an additional \$12 billion dollars are needed to fund the 180,000 manual tracers needed \cite{MCTCost}. Manual contact tracing is also susceptible to errors with the United Kingdom's Test and Trace service losing 15,000 positive COVID-19 cases between September 25 and October 2 \cite{UKBlunder}.

Digital contact tracing is a relatively new method of fighting pandemics. Although the Bluetooth technology for digital contact tracing was first validated in 2014 \cite{BluetoothTech}, contact tracing apps have only recently been implemented to fight the COVID-19 pandemic. Since digital contact tracing apps require a critical mass of a population in order to be effective, a key goal is to convince enough people to use the app. In \cite{targetcommunity}, it is suggested that targeting small communities like universities first will allow the app to be used by enough people within that community. This is more feasible than expecting a significant proportion of the population at large to use the app. Many universities have experienced outbreaks \cite{nytuniv} and are looking for ways to prevent further spread. Currently, Georgia Tech, Carnegie Mellon, Grand Valley State University, and even the city of Santa Fe are beginning to adopt contact tracing apps like NOVID \cite{novid}. Currently, there are not enough users within these communities to prove the effectiveness of contact tracing apps. We will use simulation to evaluate the effectiveness of digital contact tracing.

Efforts to model diseases date back to the 1920s with the creation of the SIR (Susceptible, Infectious, Recovered) model \cite{OGSIR} where people in a fully mixed population are modeled as being Susceptible, Infectious, or Recovered. The SEIR (Susceptible, Infectious, Exposed, Recovered) model \cite{SEIR} is a variant of the SIR framework used for diseases with longer incubation periods. It includes the Exposed stage which are infected individuals who do not show symptoms. More recently, network based models, where humans are vertices and contacts are edges, have been adopted. Epidemiological models have been instrumental in encouraging preventative measures in the COVID-19 pandemic such as masking \cite{dekaiMaskPaper,maskPaper1}, social distancing \cite{distIndia,TestandDistanceLockdownPaper}, testing \cite{screenTesting,testingNotLockdown}. In particular, \cite{dekaiMaskPaper,screenTesting} are network based models.

Since digital contact tracing relies on the exact contacts that occur in a population, it is best modeled with a contact network where people are vertices and contacts are edges. Recent efforts to simulate digital contact tracing \cite{NHSXReport,Dig2} have been done using synthetic networks where households and communities are constructed with random processes such as full mixing, and each pair of people have the same probability of contact. However, very little is known about the exact structure of human contacts and how these structures affect digital contact tracing. Additionally, those models assume perfect COVID-19 tests which changes optimal strategies of digital contact tracing and testing.


This paper presents an enhanced network based SEIR model of the COVID-19 pandemic with digital contact tracing and testing strategies that incorporates variations in infectivity and test sensitivity. In contrast to previous work on digital contact tracing, the networks in the model are generated from a real world data set of interactions among 180 students of a high school in France \cite{multiDayData}. Since the data set was only recorded over 7 days, we use a MUNGE-like heuristic to generate additional days for the model. We present a new method to extend temporal weighted graphs in order to perform simulations on a larger population of 5000 people.

Our model incorporates test sensitivity and shows that it has a significant impact on digital contact tracing strategies. The model simulates a new digital contact tracing strategy developed by NOVID \cite{novid} called the pre-exposure notification system. The pre-exposure notification system acts like a "social radar" telling app users how close they are to COVID-19, allowing them to take extra precautionary measures. In contrast with traditional SEIR models, test sensitivity is modeled to change over time depending on the time of symptom onset \cite{testSensitivity}. The incubation period is sampled from the COVID-19 incubation period distribution \cite{incubation}. The infectiousness of an individual is modeled to change over the infectious period and is fitted to the function in \cite{ViralShedding}.

The model shows that the traditional strategy of quarantining direct contacts reduces infections by less than 10\% when more than half the population is asymptomatic. Testing second and third degree contacts reduces infections by up to 40\% when 70\% of the population uses the app. The pre-exposure notification system reduces infections by an additional 43\% and reduces the number of quarantines required by 51\%. Quarantining second degree contacts reduces infections but leads to a high number of quarantines. If large proportions of the population are asymptomatic, periodic testing reduces infections by an additional 41\%. However, periodic testing without tracing reduces infections by only 3\%. The most effective strategy discussed in this work was combined the pre-exposure notification system with testing second and third degree contacts. This strategy reduces infections by 18.3\% when 30\% of the population uses the app, 45.2\% when 50\% of the population uses the app, 72.1\% when 70\% of the population uses the app, and 86.8\% when 95\% of the population uses the app. When simulating the model on an extended network of 5000 students, the results are very similar with the contact tracing app reducing infections by up to 79\%.

\subsection{Paper Outline}

In Section \ref{network}, we present the contact network generation process. Section \ref{covid} outlines the SEIR based model of COVID-19 spread. In Section \ref{app}, we create the model of the contact tracing app. In Section \ref{results} we present the results of 8 simulated scenarios. In Section \ref{S1}, the traditional strategy of quarantine of first degree contacts is investigated. In Section \ref{S2}, testing of second and third degree contacts is incorporated. In section \ref{S6}, the pre-exposure notification system is investigated. Section \ref{S7} investigates periodic testing. Section \ref{S8} simulates the model on an extended version of the graph. In Section \ref{econ}, we estimate the economic value of the contact tracing app. In Section \ref{conclusion}, we provide a summary and concluding remarks.

\section{Model of Contact Network}\label{network}

The first component of the model is the contact network which encapsulates the interactions between individuals in the simulation. The model iterates through discrete timestamps each representing a day in the simulation. Each day in the model, individuals come into contact, and these contacts ultimately determine how the virus transmits. The contact network on day $t$, $G_t$ is defined to be a graph where each vertex $v_1,...,v_n$ represent an individual in the model and the edge weight $e_{ij}$ between $v_i$ and $v_j$ represents the total amount of time person $i$ and person $j$ have spent in contact with each other during this time step. We do not distinguish between times of the day of the contact or number of contacts between the same individuals. This is in accordance with a recent CDC policy change that defines a close contact to be measured using cumulative contact time over the course of a day \cite{CDCAppendix}.


Traditional SEIR models assume uniform mixing where each individual has the same probability of coming into contact with every other individual. Since digital contact tracing involves the exact contacts that occur in a population, a realistic contact network is needed. Temporal networks in the model are generated using a publicly available high-resolution data set \cite{multiDayData} which used RFID (Radio Frequency Identification Devices) to record all contacts between 180 students from Lyc\'ee Thiers high school in France over the course of 7 days.

Since the simulation is over longer periods of time, we present a heuristic for generating additional days. Define $H_a$ to be the contact network in day $a$ of the data set where $1 \le a \le 7$. The edge weight $H_{a,i,j}$ is the duration of contact between people $i$,$j$ on day $a$ from the data set. The heuristic is similar to the MUNGE algorithm \cite{MUNGE} which generates synthetic training data. The algorithm picks an initial data value, and for each feature, with a certain probability, swaps that feature with the feature of its nearest neighbor. In this heuristic, we use a random day rather than the nearest neighbor. Initially, 2 distinct days $a$,$b$ are chosen at random from the data set. Starting with the contact graph for day $a$, for the contact duration between $v_i$ and $v_j$ with probability $0.5$, we replace that contact duration with $H_{b,i,j}$. On average, the generated contact network $G_t$ has half of its edges equal to the corresponding edges in $H_a$, and the other half equal to $H_b$.


To simulate the app on larger networks, we present a method to generate larger networks from the original data set. We create a modified version of the Albert-Barab\'asi process that is adjusted for constructing weighted temporal graphs and maintains the average degree of the vertices. 


The principal idea of the Albert-Barab\'asi process is that nodes of high degree are more likely to interact with new nodes. Given a temporal graph $G$ with $n$ vertices where the graph on timestamp $t$ is $G_t$, extra people are added one at a time with the following process:

A new vertex $v$ is added, and for each existing $w$, a random third vertex $u$ is chosen. The temporal edge weight between $v$ and $w$ is determined to be the same as the edge weight between $w$ and $u$: $G_{t,v,w}=G_{t,u,w}$ for all $t$. This step is the same as in Albert-Barab\'asi except adjusted to fit a weighted temporal graph. To preserve the average degree of the vertices, each of the original edges of $G$ are deleted with probability $\frac{1}{n-1}$. Note that the generated graph no longer has the scale-free property.

The extended network is initialized to be the original graph in the data set and extended in accordance to the above procedure to a total of 5000 individuals.

\section{Model of COVID-19 Spread}\label{covid}

The SEIR framework organizes the people in the simulation into one of the following four states: Susceptible (S), Exposed (E), Infectious (I), or Recovered (R). Individuals in the Susceptible stage have not been infected and therefore are susceptible to the virus. Exposed individuals have been infected with the virus but do not yet show symptoms. The Infectious stage begins when infected individuals show symptoms. Finally, the Recovered stage contains individuals who have recovered or otherwise removed from the model and are now immune from further spread. Individuals in both the Exposed and Infectious stage can infect others. An individual moves onto the next stage according to the following rules:

\begin{itemize}

\item $S\rightarrow E$: Individuals in the susceptible stage can only be exposed if they come into contact with an infected individual. Infections will be discussed in later sections.

\item $E\rightarrow I$: After being exposed to a virus, the period of time before symptom onset is called the Incubation period, typically within 2 to 14 days. According to \cite{incubation}, the distribution of incubation periods is approximately the log-normal distribution with parameters $\mu=1.621$, $\sigma=0.418$. The log-normal distribution is defined as follows: Let $Z$ be a random variable with normal distribution with mean $\mu$ and deviation $\sigma$, then $X=e^{Z}$ where $X$ is the random variable with log-normal distribution with parameters $\mu,\sigma$. To determine the incubation period of an individual, we take a random sample from this distribution rounded to the nearest positive integer.

\item $I\rightarrow R$:Individuals with moderate symptoms stop being infectious around 10 days from symptom onset \cite{CovidLength}. As in traditional SEIR models, we assume that every time stamp after symptom onset, there is a probability $\lambda=0.11$ which is $\frac{1}{9}$ that an Infectious person recovers or is removed from the model.

\end{itemize}

As in traditional SEIR models, the functions $S(t),E(t),I(t),R(t)$ are the number of individuals in the compartments susceptible, exposed, infections, and recovered respectively at time $t$. We defined $Q(t)$ to be the number of individuals quarantined at time $t$ but have not received a positive test result and $T(t)$ to be the number of individuals who have received a positive before or during time $t$ but have not recovered. Individuals counted in $T(t)$ are in quarantine as well. Quarantines will be discussed in Section \ref{app}. In the rest of the paper, for each individual $v_i$, we will refer to $E_{v_i}$, $I_{v_i}$, and $R_{v_i}$ as the time of exposure, symptom onset, and recovery respectively of $v_i$. Additionally, $T_{v_i}$ is the time of the first positive test of $v_i$.

In realistic scenarios, individuals can infect others before symptom onset with the level of infectiousness changing throughout of the infection. The relative infectiousness of an individual is calculated in \cite{ViralShedding}. We will call this function $ID(t)$ where the input $t$ is the number of days since symptom onset. $ID(t)$ is taken to be the function in \cite{ViralShedding} where infectiousness was assumed to start 5 days before symptom onset. Individuals are shown to be more infectious before symptom onset than after. We assume the infectiousness of an individual, which is defined to be the probability of infection given a 20-second contact, is some constant multiple, $p$, of this function.

Consider the graph $G_t$ to be the contact network during timestamp $t$ in the model, with vertices $v_1,...,v_n$ representing people and edge weight $e_{ij}$ to be the total contact duration between $v_i,v_j$ during day $t$. Assuming infection to be an event that can happen during the course of a contact, if $v_i$ is in either the exposed or infectious compartments, and $v_j$ is susceptible, we model the probability that $v_j$ becomes exposed to be
$$1-(1-p ID(t-I_{v_i}))^d$$
where $t-I_{v_i}$ is the number of days since symptom onset of $v_i$.

The basic reproduction number, or $R_0$, is the expected number of secondary infections caused by a single infectious individual. Estimates of the value of $R_0$ range from $1.5$ to $6.7$ with the median value being $2.8$ \cite{R0Estimate} and the methods used to estimate $R_0$ varies widely between studies \cite{R0Conundrum}. Through simulation, we calculate the value of $R_0$ as a function of $p$ as shown in Figure \ref{pR0}. The value of $p$ is ranged in increments of $0.0025$ from $0$ to $0.225$ for a total of 100 values. For each value of $p$, we run the simulation 1,800 times where each individual is the seed infection 10 times and the $R_0$ value is measured as the average number of secondary infections caused by the seed infection. As shown in Figure \ref{pR0}, the scenario when $p=0.10$ yields an $R_0$ value of $2.8$. In the rest of this work, the value of $p$ will be set to $0.10$

\begin{figure}[H]
    \centering
    \includegraphics[width=8cm]{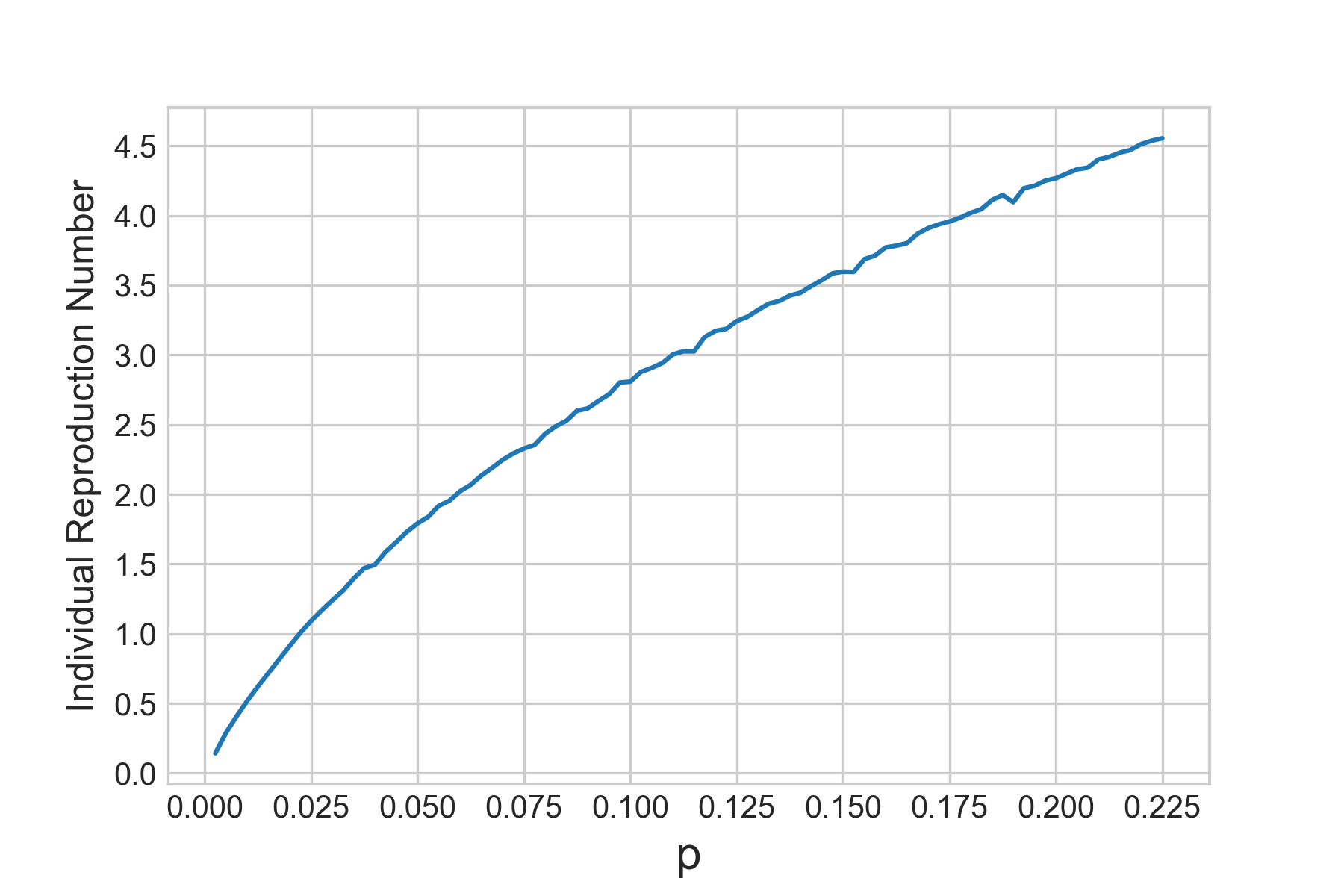}
    \caption{Probability of infection vs individual reproduction number}
\end{figure}\label{pR0}

According to \cite{CDCScenarios}, 40\% of the population remain asymptomatic throughout their infectious periods. Again, estimates vary widely and it is noted that this value could range from 10\% to 70\%. In \cite{80Kids}, it is estimated that over 80\% of young individuals are asymptomatic. The possibilities of when 20\%, 40\%, 60\%, and 80\% are asymptomatic are tested in the results. Asymptomatic individuals have similar viral loads to symptomatic individuals \cite{AsympInf}. Thus, we will assume that asymptomatic individuals are as infectious as symptomatic individuals.

Testing is a key component of contact tracing. The contact tracing app relies on positive tests to identify cases of COVID-19. We adjust test sensitivities based on days from symptom onset to the distribution calculated in \cite{testSensitivity}. The false-negative rate of RT-PCR COVID-19 tests are between $67\%$ to $100\%$ before symptom onset, and fall to $20\%$ to $40\%$ after symptom onset \cite{testSensitivity}. The median number of days from symptom onset to taking a test is 3 days with interquartile range 1 to 6 \cite{CDCScenarios}. We approximate this distribution by assuming that each day there is a probability of 19\% that symptomatic individuals are tested. This preserves the median of 3 days and interquartile range 1 to 6. We do not account for false-positive tests or individuals symptomatic with a disease unrelated to COVID-19. Test results are received after a delay of 1 day. This is similar to the delay on university campuses \cite{UMichTest}. After receiving a positive COVID-19 test result, we assume the person will remain in quarantine until recovery.

\section{Model of Contact Tracing App}\label{app}

The contact tracing app has incomplete information about the contact network and the states of the individuals. Ultrasound apps such as NOVID can measure distances to the resolution of inches and detect contacts with accuracy over 99.6\% \cite{novid}. We will assume all contacts between individuals with the app are sensed. Thus, the contact tracing app can detect the subgraph of the contact map induced by the set of app users. Contact tracing apps support the consideration of degree $k$ contacts rather than just direct contacts. This is not possible with manual contact tracing. A contact is defined to be a tuple $(p_1,p_2,t,d)$ that contains the following four pieces of information: $p_1,p_2$ are the people involved in the contact, $t$ is the day of contact, and $d$ is the duration of the contact measured in units of 20 seconds. Since the digital contact tracing app cannot determine the time of symptom onset, the estimated transmission probability of the contact is $(1-(1-p')^{d_i})$. The value $p'=\widehat{ID} p$ is substituted for $pID(t)$ where $\widehat{ID}=0.11$ is the average value of $ID(t)$ during the 6 most infectious days. In addition to contacts, users report all positive test results. If a user has tested positive, we assume that this user will report recovery. In the contact tracing app, we define a degree $k$ contact to be a sequence of contacts $c_1,…,c_k$ that satisfy the following properties:
\begin{enumerate}

\item The individuals $p_1,...,p_{i+1}$ involved in the contacts form a chain: $c_i=(p_i,p_{i+1},t_i,d_i)$.

\item $p_1$ has reported a positive test: $t_1 \le T_{p_1} \le t_1+10$. Through simulation of 1800 outbreaks without interventions, 95\% of all symptomatic individuals took a test within 8 days of symptom onset. Since infectiousness becomes significant 2 days before symptom onset, the contacts from 10 days before the test must be recalled.

\item $p_i,p_{i+1}$ have not reported recovery and thus can transmit or catch the virus: $R_{p_i},R_{p_{i+1}}>t_i$ for each $i$.

\item The serial interval is between 1 and 10 days: $t_i+1\le t_{i+1}\le  t_i+10$. The serial interval is defined to be the duration from the exposure time of the infector($t_i$) to the exposure time of the infected($t_{i+1}$). Through simulation of the model 1800 time, 94\% of all serial intervals are within 10 days.

\end{enumerate}

The weight of this contact chain is the product of the estimated transmission probabilities of each contact:
$$\prod_{i=1}^k (1-(1-p')^{d_i}).$$

First and second degree contacts are computed using contact data from previous days directly after $p_1$ reports a positive test.  Larger degree contacts are computed recursively. For each person $v_i$ in the model, we keep track of a matrix $M_i$ where $M_{i,l,t}$ is the total sum of contact chains of length $l$ that affect person $v_i$ at time $t$. Using the contacts, the app can calculate $M_i$ recursively.

On day $t$, if $v_i$ has contacts with $v_{a_1},...,v_{a_k}$ with durations $d_1,...,d_k$ respectively, where $v_i,v_{a_1},...,v_{a_k}$ have not reported recovery by day $t$, then:

$$M_{i,l,t}=\sum_{s=1}^k \left( \left(1-(1-p')^{d_s}\right)\sum_{r=1}^{10} M_{s,l-1,t-r} \right).$$

For all of $v_i$'s contacts, we take the sum of all contact chains of length $l-1$ and multiply by the estimated transmission probability $1-(1-p')^{d_s}$ to obtain the sum of all contact chains of length $l$.

A person $x$ is a degree $k$ contact on day $t$ if the sum of all weights of their contacts with degree $\le k$ is above $10\%$:

$$\sum_{i=1}^k\sum_{j=1}^{10} M_{x,i,t-j} \ge C.$$

The contact cutoff value, $C$, only affects the quarantine rules and not simulating the disease transmission. Essentially, direct contacts that are calculated to be infected with probability at least $C$ are quarantined. The value of $C$ is set to $10\%$ by default. 
By default, first degree contacts are quarantined.

The pre-exposure notification system, as implemented in NOVID \cite{novid}, acts like a "social radar" telling users how close they are to COVID-19 by showing the number of positive cases at each distance in their social network. App users can see the number of neighbors of distance $d$ that are COVID-19 positive. We assume that first, second, and third degree neighbors will take precautionary measures. For second degree contacts, we assume a 75\% reduction in contacts. For third degree contacts, we assume a 50\% reduction in contacts.

\section{Results}\label{results}

At the beginning of the simulation, exactly one individual is exposed while the rest are susceptible. The simulation runs for 120 days. There are 180 individuals and for each simulation we run 1800 trials where each individual starts as the seed infection 10 times. We simulate the cases when $0\%$, $30\%$, $50\%$, $70\%$, and $95\%$ of the population use the app.


We simulate the following 5 scenarios:
\begin{enumerate}
\item Quarantine of first degree contacts
\item Quarantine of first degree contacts with followup testing of second and third degree contacts
\item Pre-exposure notification system
\item Pre-exposure notification system with periodic testing
\item Scenario 4 on the extended graph
\end{enumerate}

We will measure the contact tracing app and COVID-19 testing configuration by 3 metrics: Total infected, total days spent in quarantine, and total tests used. In the tables below, the App Proportion column shows the proportion of individuals in the model that use the app. The Asymptomatic column show the proportion of individuals that are asymptomatic. The Infected column shows the average number of individuals infected after 120 days. Quarantines are split into 3 categories: false, true, tested. False quarantines are quarantined individuals who do not have COVID-19. Individuals in the True quarantine category are infected with COVID-19 but have not received a positive test result. Tested quarantines are those who have tested positive. We distinguish the Tested quarantines because these infections have been confirmed by test while True and False quarantines are predicted by the quarantine rules. Each of the quarantine columns show the total number of days spent in quarantine across all individuals over 120 days. The Tests Used column shows the number of tests used in the simulation after 120 days.

In scenario 1, the most basic strategy of quarantining first degree contacts is not effective at high asymptomatic levels, suggesting that a more comprehensive testing strategy is required. Scenario 2 shows that testing second and third degree contacts greatly increases the effectiveness of the app. The pre-exposure notification system, which warns second and third degree contacts to take extra precautionary measures, can reduce infections by an additional 40\% while also reducing the number of quarantines by up to 50\%. IN Scenario 4, it is shown that periodic testing without contact tracing reduces infections by less than 3\%. Each app user decreases the economic cost of COVID-19 by \$2,841 at 50\% app usage and \$4,185 at 70\% app usage.


\subsection{Scenario 1 --- Quarantine of first degree contacts (Traditional Strategy)}\label{S1}

This is the most basic strategy where only first degree contacts are quarantined. Every day, symptomatic individuals have a 19\%  chance of getting a test as discussed in Section \ref{covid}. The simulation is performed on the original network of 180 students.

\begin{table}[H]
    \caption{Quarantine direct contacts, no followup testing}
    \centering
    \begin{tabular}{|c|c|c|c|c|c|c|}
    \hline
    \multicolumn{2}{|c|}{Situation} & \multirow{ 2}{*}{Infected} & \multicolumn{3}{c|}{Quarantine} & \multirow{ 2}{*}{Tests Used}\\ 
    \cline{1-2} \cline{4-6}
    App Proportion & Asymptomatic &&False&True&Tested&\\
    \hline
    
    0\%&40\%&106.05 & 0 & 0 & 291.87 & 53.57  \\
    30\%&40\%&101.20 & 78.78 & 39.73 & 278.42 & 50.97\\
    50\%&40\%&97.15 & 195.71 & 94.06 & 265.96 & 48.97\\
    70\%&40\%& 87.48 & 329.17 & 149.91 & 242.71 & 44.41 \\
    95\%&40\%& 70.25 & 477.30 & 200.56 & 194.77 & 35.39\\
    \hline
    0\%&80\%& 110.43 & 0 & 0 & 101.43 & 18.43 \\
    30\%&80\%&107.99 & 33.53 & 19.23 & 98.83 & 18.09\\
    50\%&80\%&108.12 & 88.47 & 52.20 & 97.12 & 17.96\\
    70\%&80\%& 101.82 & 159.04 & 91.53 & 94.11 & 17.04 \\
    95\%&80\%& 99.73 & 282.78 & 156.70 & 90.58 & 16.76\\
    \hline
    \end{tabular}
    \label{S1R2}
\end{table}

\begin{figure}[H]
  \centering
  \subfloat[Total infected.]{\includegraphics[width=8.5cm]{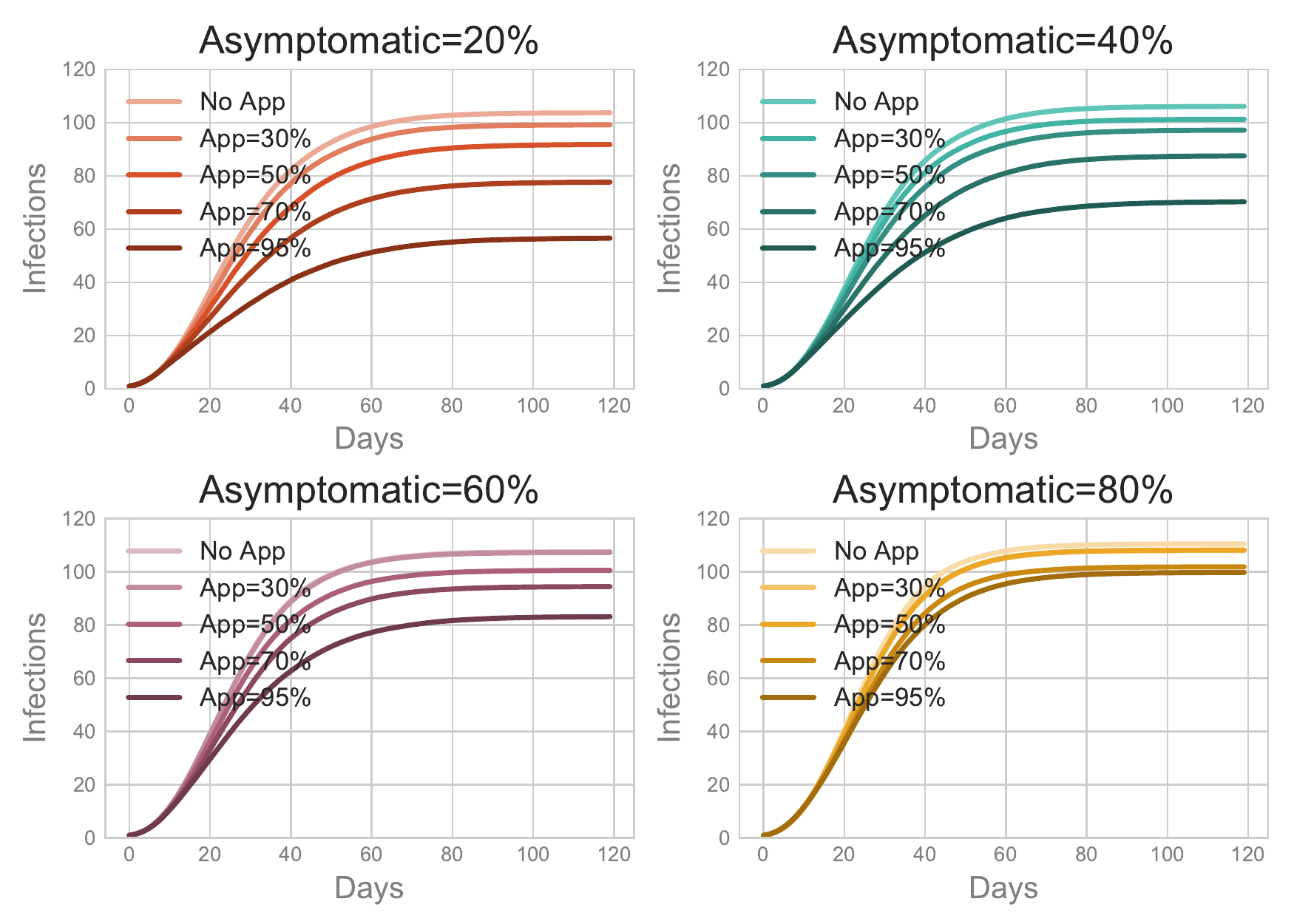}\label{S1R2Line}}
  \subfloat[Total infected but not recovered or tested.]{\includegraphics[width=8.5cm]{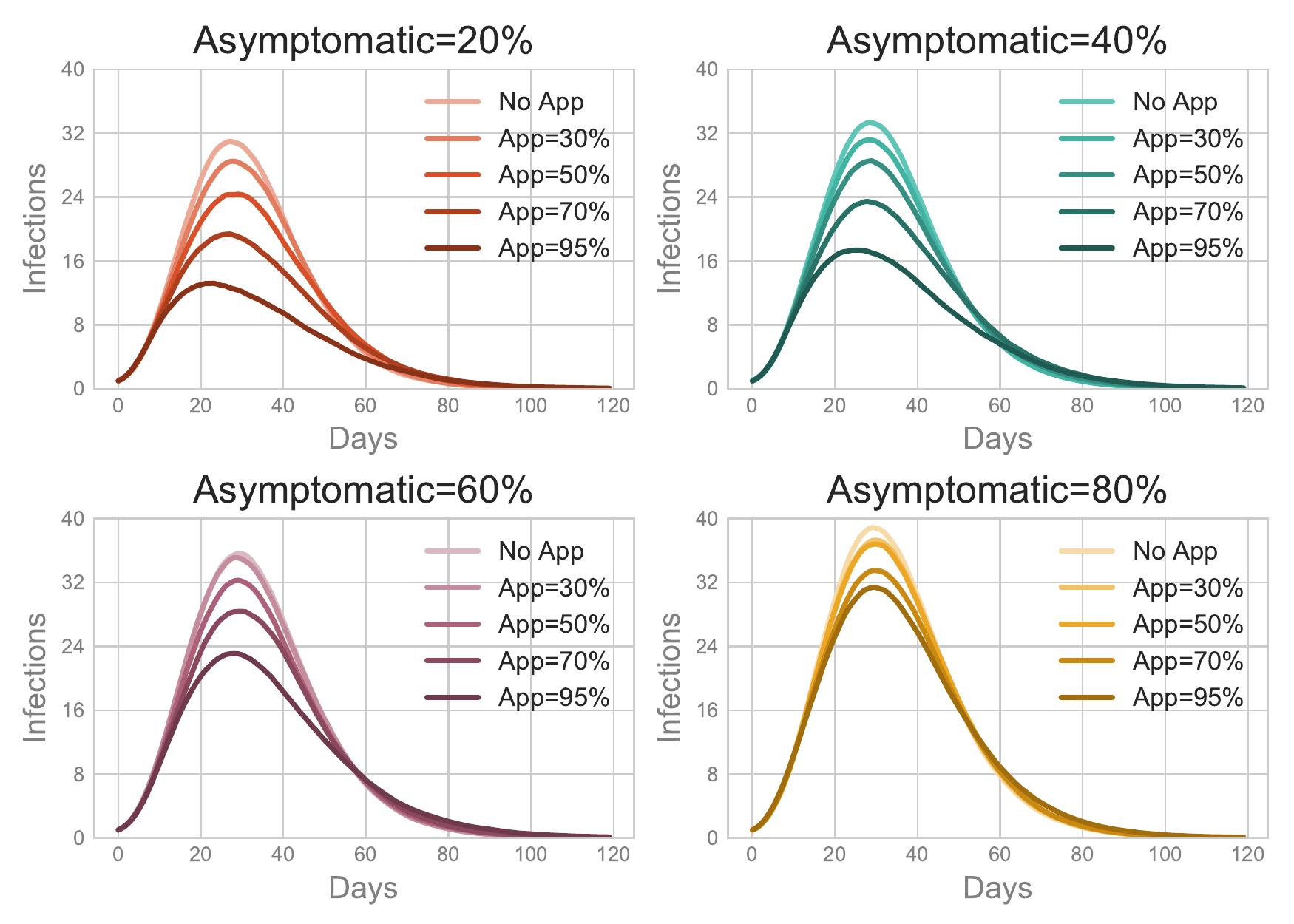}\label{S1R2Inf}}
  \caption{Quarantine direct contacts, no followup testing}
\end{figure}

From Table \ref{S1R2}, we can find out that the effectiveness of the app depends on the percentage of the people use the app. At 40\% asymptomatic ratio, the contact tracing app reduces infections by 8.4\% at 50\% app usage and 17.5\% at 70\% app usage. The effectiveness of the app increases doubles from adding the extra 20\% of app users. This shows that it is very important to have a majority of app users. 

A limitation of this strategy is that it is ineffective at high asymptomatic ratios. At 80\% asymptomatic ratio the contact tracing app reduces infections by less than 10\% in all cases. This is because in this scenario, since COVID-19 testing relies on symptomatic individuals, the effectiveness of the app is greatly reduced at high asymptomatic levels. Less than 20 tests are used while over 100 students are infected which indicates that the majority of infections are undetected. The results in this scenario suggest that more comprehensive testing strategies are needed, especially when large proportions of the population are asymptomatic.

The cost of using the app is that quarantines increase as more people use the app. In all cases, the number of false quarantines is approximately double the number of true quarantines. The time of peak infections is similar for each app usage. In Figure \ref{S1R2}, infections rise dramatically, peaking at around 30 days after the initial infection and then falls as herd immunity is reached.  Because quarantine rules require a history of contacts to be recalled, the initial 10 days show no difference between the graphs of each app usage level. At 60\% and 80\% asymptomatic ratio, the app is unable to reduce infections significantly.

\subsection{Scenario 2 --- Testing of second and third degree contacts}\label{S2}

In this scenario, first, second, and third degree contacts are tested every 3 days. All other parameters are the same as in scenario 1. The simulation is performed on the original network of 180 students.

\begin{table}[h]
    \caption{Quarantine direct contacts, test second, third degree contacts}
    \centering
    \begin{tabular}{|c|c|c|c|c|c|c|}
    \hline
    \multicolumn{2}{|c|}{Situation} & \multirow{ 2}{*}{Infected} & \multicolumn{3}{c|}{Quarantine} & \multirow{ 2}{*}{Tests Used}\\ 
    \cline{1-2} \cline{4-6}
    App Proportion & Asymptomatic &&False&True&Tested&\\
    \hline
    0\%&40\%& 104.61 & 0 & 0 & 286.37 & 52.68 \\
    30\%&40\%&98.53 & 113.23 & 32.84 & 325.58 & 107.60\\
    50\%&40\%&83.02 & 259.09 & 73.70 & 340.81 & 234.49\\
    70\%&40\%& 63.33 & 395.14 & 101.05 & 315.52 & 470.46 \\
    95\%&40\%& 27.87 & 314.2 & 77.42 & 172.38 & 670.94\\
    \hline
    0\%&80\%& 111.52 & 0 & 0 & 100.72 & 18.71 \\
    30\%&80\%&106.79 & 88.73 & 23.59 & 156.34 & 62.29\\
    50\%&80\%&95.37 & 253.73 & 71.30 & 250.42 & 188.05\\
    70\%&80\%& 78.03 & 418.75 & 110.40 & 304.72 & 400.50 \\
    95\%&80\%& 44.10 & 423.31 & 105.73 & 238.87 & 645.00\\
    \hline
    \end{tabular}
    \label{S2R2}
\end{table}

\begin{figure}[H]
  \centering
  \subfloat[Total infected.]{\includegraphics[width=8.5cm]{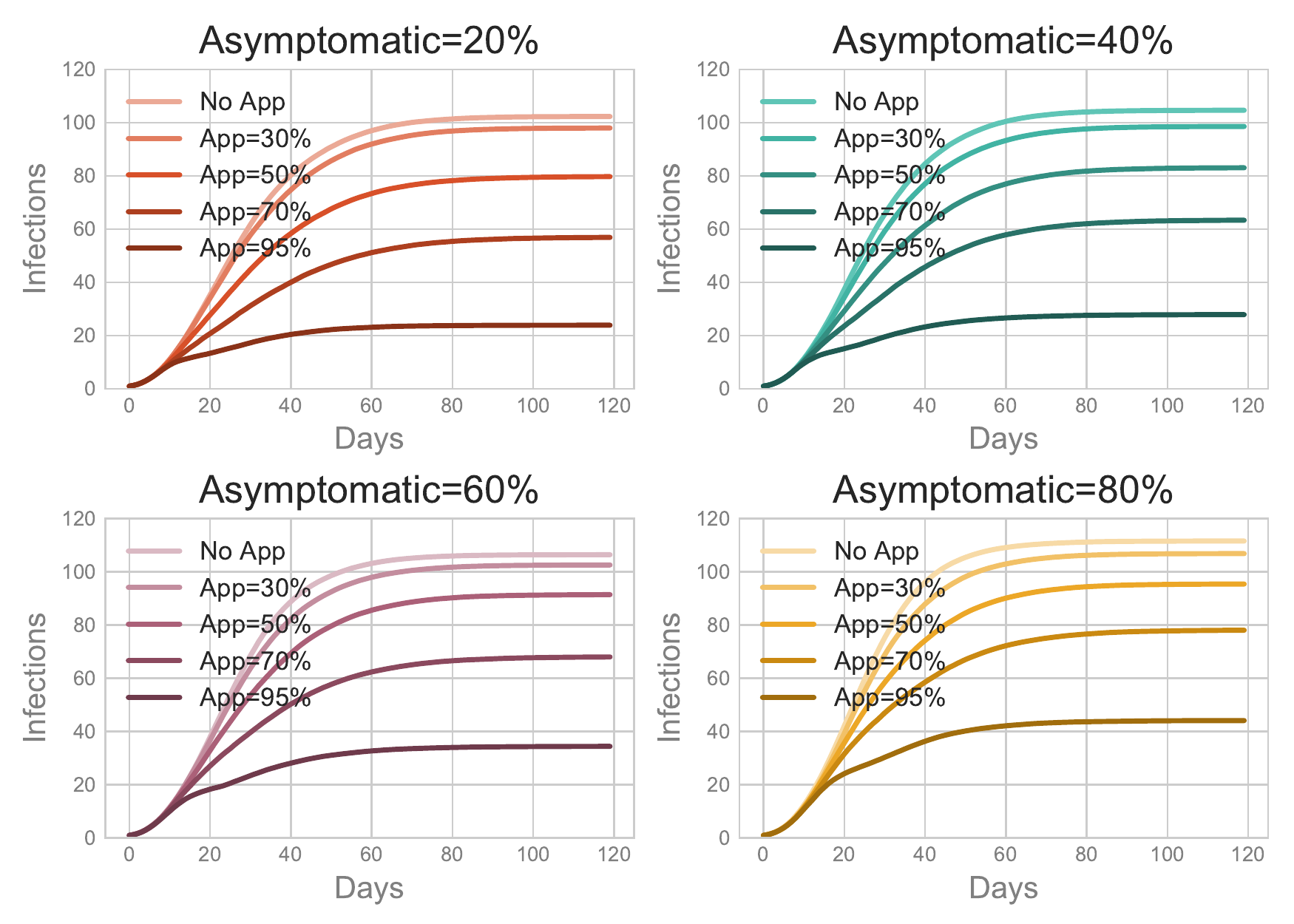}\label{S2R2Line}}
  \subfloat[Total infected but not recovered or tested.]{\includegraphics[width=8.5cm]{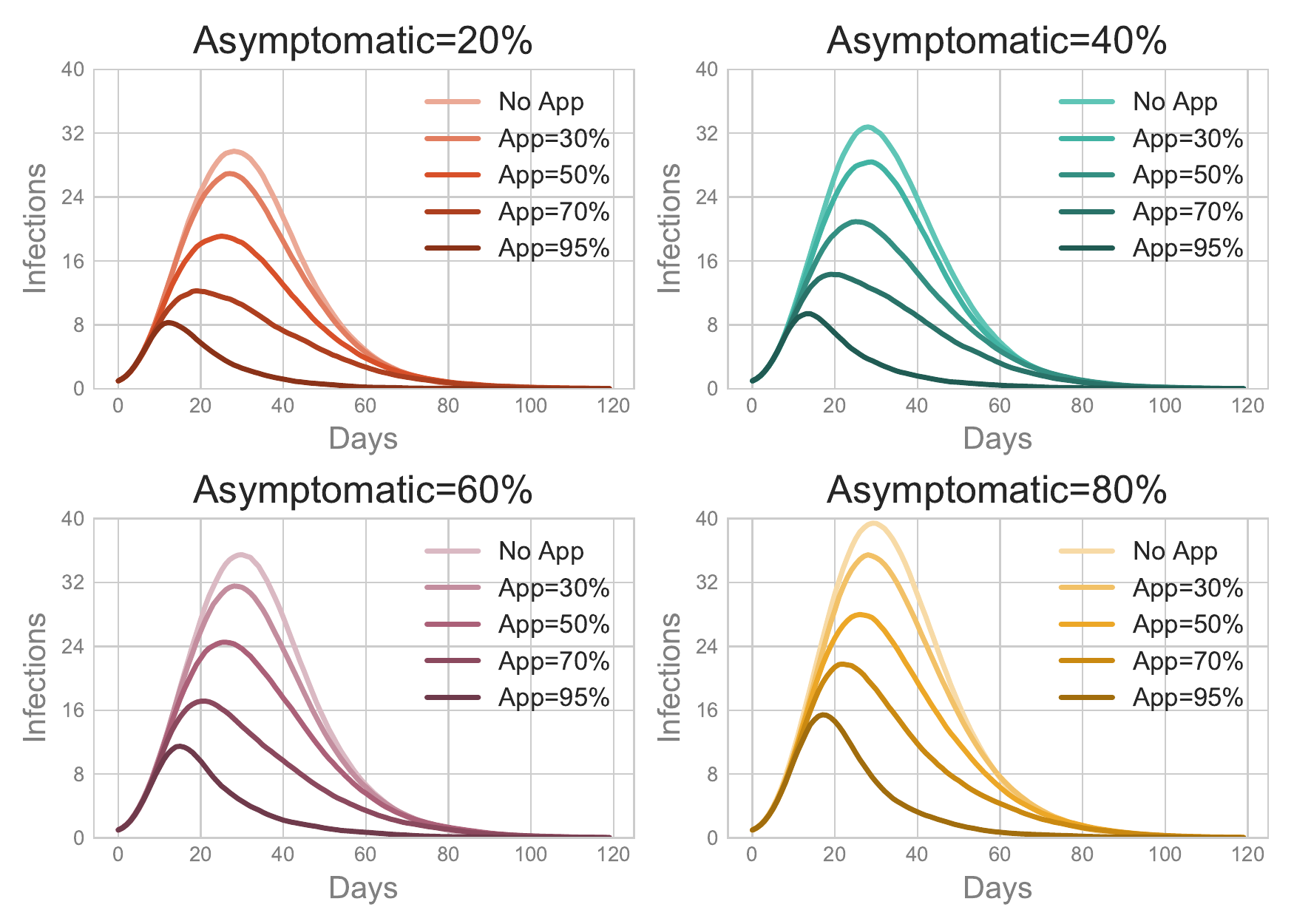}\label{S2R2Inf}}
  \caption{Test second, third degree contacts}
\end{figure}
As in scenario 1, the effectiveness of the app increases greatly as the number of app users increases. In Table \ref{S1R2}, at 40\% asymptomatic ratio, the contact tracing app reduces infections by 20.6\% at 50\% app usage and 39.5\% at 70\% app usage. These reductions are over 2 times greater than in scenario 1. Notably, at 95\% app usage, infections are reduced by 73.4\%, which is almost twice the reduction at 70\% app usage. Compared to scenario 1, the number of true quarantines decreases while the number of tested quarantines increases at 40\% asymptomatic ratio. At 80\% asymptomatic ratio, the contact tracing app reduces infections by 14.5\% at 50\% app usage and 30.0\% at 70\% app usage. Again, this strategy is significantly more effective than scenario 1. However, the number of quarantines increases significantly in all categories compared to scenario 1. In particular, the number of tested quarantines rises by 224\% at 70\% app usage. This is to be expected since the number of tests used rises significantly, from less than 20 to over 400. Again, the app is significantly more effective at 95\% app usage, with infections being reduced by 60.05\%. Although results for 40\% and 80\% asymptomatic ratio are similar for app proportions less than 70\%, at 95\% app proportion, the case of 80\% asymptomatic yields 58\% more infections.

Similar to scenario 1, the graphs are similar during the first 10 days and then diverge significantly. However, at high app usage, infections peak before 20 days while no app usage yields peak infections at around 30 days. In conclusion, testing second and third degree contacts greatly increases the effectiveness of the app especially at higher asymptomatic proportions. Thus, the increase in test usage and quarantines is justified.

\subsection{Scenario 3 --- Pre-exposure notification system}\label{S6}

In this scenario, we simulate the pre-exposure notification system. For second degree contacts, we assume a 75\% reduction in contacts. For third degree contacts, we assume a 50\% reduction in contacts. All other parameters are the same as those in scenario 2. The simulation is performed on the original network of 180 students.

\begin{table}[H]
    \caption{Quarantine direct contacts, pre-exposure notification system}
    \centering
    \begin{tabular}{|c|c|c|c|c|c|c|}
    \hline
    \multicolumn{2}{|c|}{Situation} & \multirow{ 2}{*}{Infected} & \multicolumn{3}{c|}{Quarantine} & \multirow{ 2}{*}{Tests Used}\\ 
    \cline{1-2} \cline{4-6}
    App Proportion & Asymptomatic &&False&True&Tested&\\
    \hline
    0\%&40\%&106.99 & 0 & 0 & 295.84 & 54.24\\
    30\%&40\%&93.48 & 92.80 & 27.40 & 303.44 & 101.73\\
    50\%&40\%&67.25 & 162.16 & 44.36 & 254.02 & 161.40\\
    70\%&40\%&36.21 & 185.80 & 45.24 & 162.85 & 213.24\\
    95\%&40\%&16.61 & 190.22 & 41.60 & 98.10 & 279.80\\
    \hline
    0\%&80\%&110.90 & 0 & 0 & 101.08 & 18.69\\
    30\%&80\%&103.19 & 87.01 & 22.46 & 150.39 & 64.83\\
    50\%&80\%&81.68 & 187.45 & 48.33 & 189.58 & 142.75\\
    70\%&80\%&54.06 & 251.14 & 61.42 & 179.84 & 220.68\\
    95\%&80\%&29.99 & 287.72 & 65.62 & 156.12 & 309.38\\
    \hline
    \end{tabular}
    \label{S6R2}
\end{table}

As in the previous scenarios, the effectiveness of the app increases as the number of app users increases. In Table \ref{S6R2}, at 40\% asymptomatic ratio, the contact tracing app reduces infections by 37.1\% at 50\% app usage and 66.1\% at 70\% app usage. At 95\% app usage, infections are reduced by 84.5\% which is an additional 40.2\% reduction compared to scenario 2. Compared to scenario 2, the total number of quarantines is reduced by 51.4\% at 70\% app usage. There is a 54.7\% reduction in the number of tests used which is caused by the reduced interactions for second and third degree contacts. This strategy is more effective even though the number of true and tested quarantines are reduced by 55.2\% and 48.4\% respectively.

At 80\% asymptomatic ratio, the contact tracing app reduces infections by 14.5\% at 50\% app usage and 30.0\% at 70\% app usage. Again, this strategy is significantly more effective than scenario 1. At 95\% app usage, the 80\% asymptomatic case yields 80.6\% more infections than the 40\% case. This is an even more pronounced gap than in scenario 2.

\begin{figure}[H]
  \centering
  \subfloat[Total infected.]{\includegraphics[width=8.5cm]{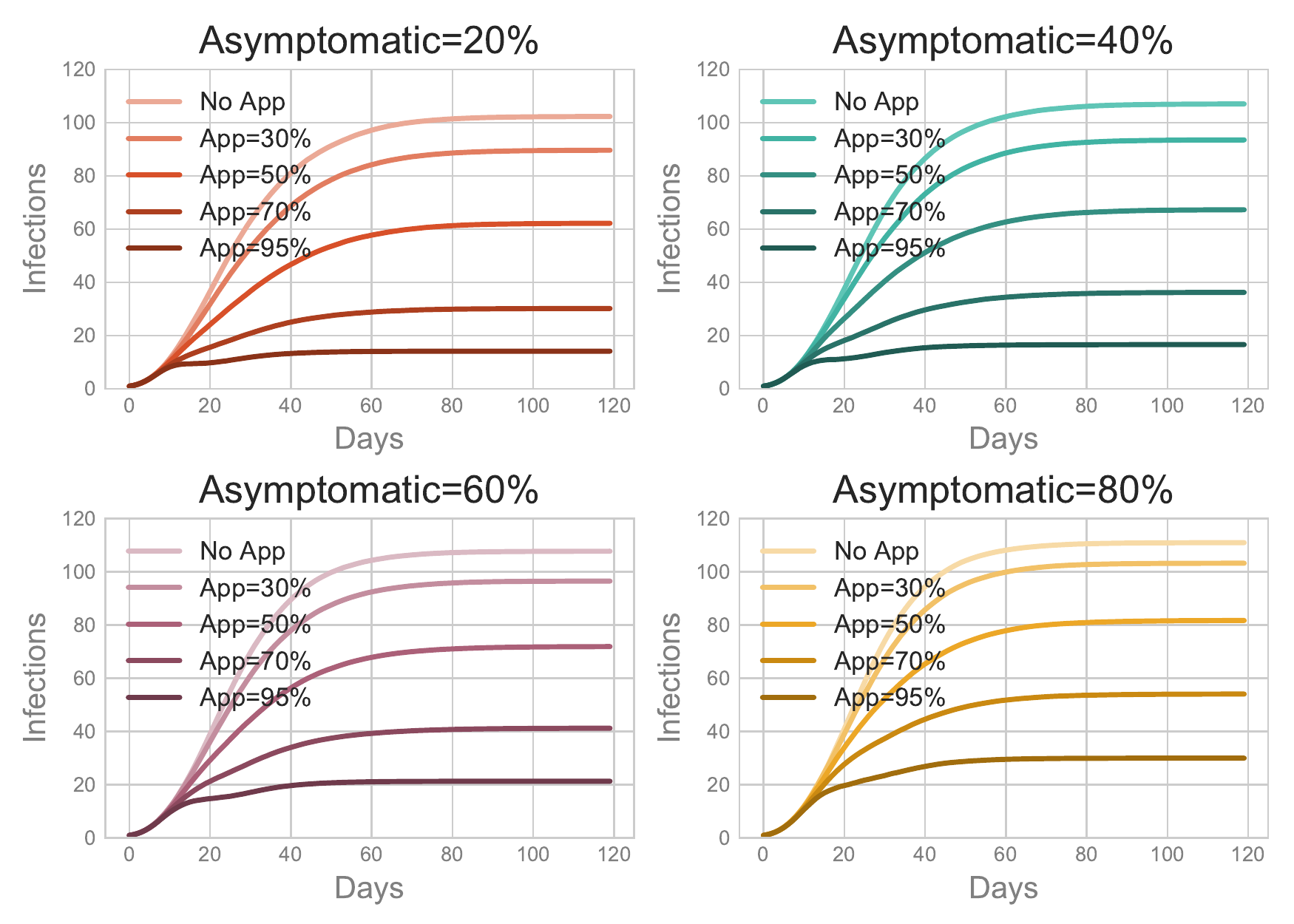}\label{S6R2Line}}
  \subfloat[Total infected but not recovered or tested.]{\includegraphics[width=8.5cm]{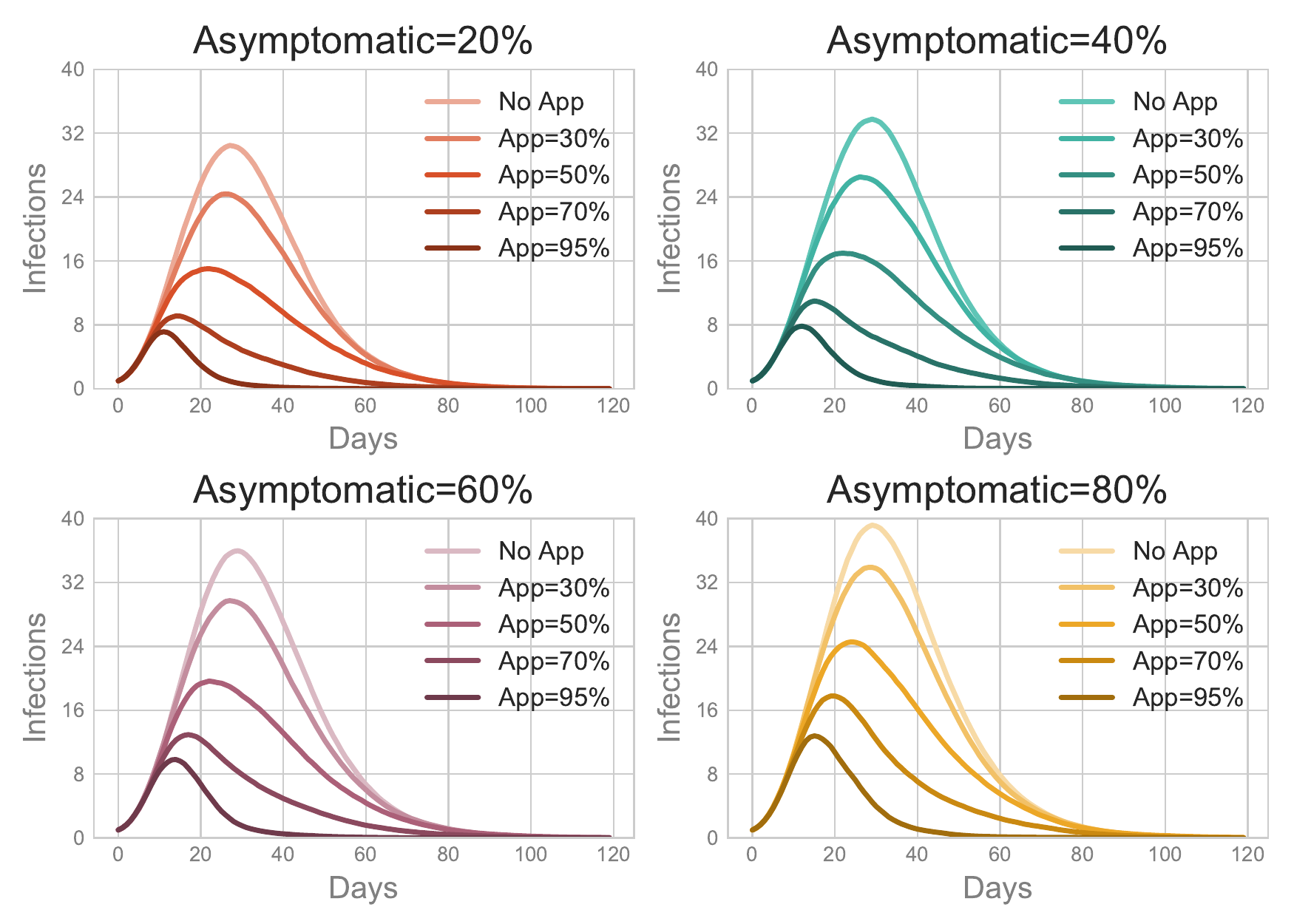}\label{S6R2Inf}}
  \caption{Quarantine direct contacts, pre-exposure notification system}
\end{figure}

\subsection{Scenario 4 --- Pre-exposure notification system with periodic testing}\label{S7}

In scenario 4, we simulate the Pre-Exposure Notification System with periodic testing every 14 days is investigated. As in scenario 3, second and third degree contacts take extra precautionary measures. Individuals in the model are tested every 14 days. The simulation is performed on the original network of 180 students.

At 40\% asymptomatic ratio, the contact tracing app reduces infections by 45.2\% at 50\% app usage and 72.1\% at 70\% app usage. At 95\% app usage, infections are reduced by 86.8\%. Scenario 3 and 4 show similar results at higher app usage. In this scenario, the app is effective even at 30\% app usage. The 30\% app usage case reduces infections by 18.1\% which is much more effective than in previous scenarios. Additionally, results are similar at all asymptomatic ratios. At 80\% asymptomatic ratio, the contact tracing app reduces infections by 37.8\% at 50\% app usage and 66.4\% at 70\% app usage. Thus, periodic testing is especially helpful when larger proportions of individuals are asymptomatic.

\begin{figure}[H]
  \centering
  \subfloat[Total infected.]{\includegraphics[width=8.5cm]{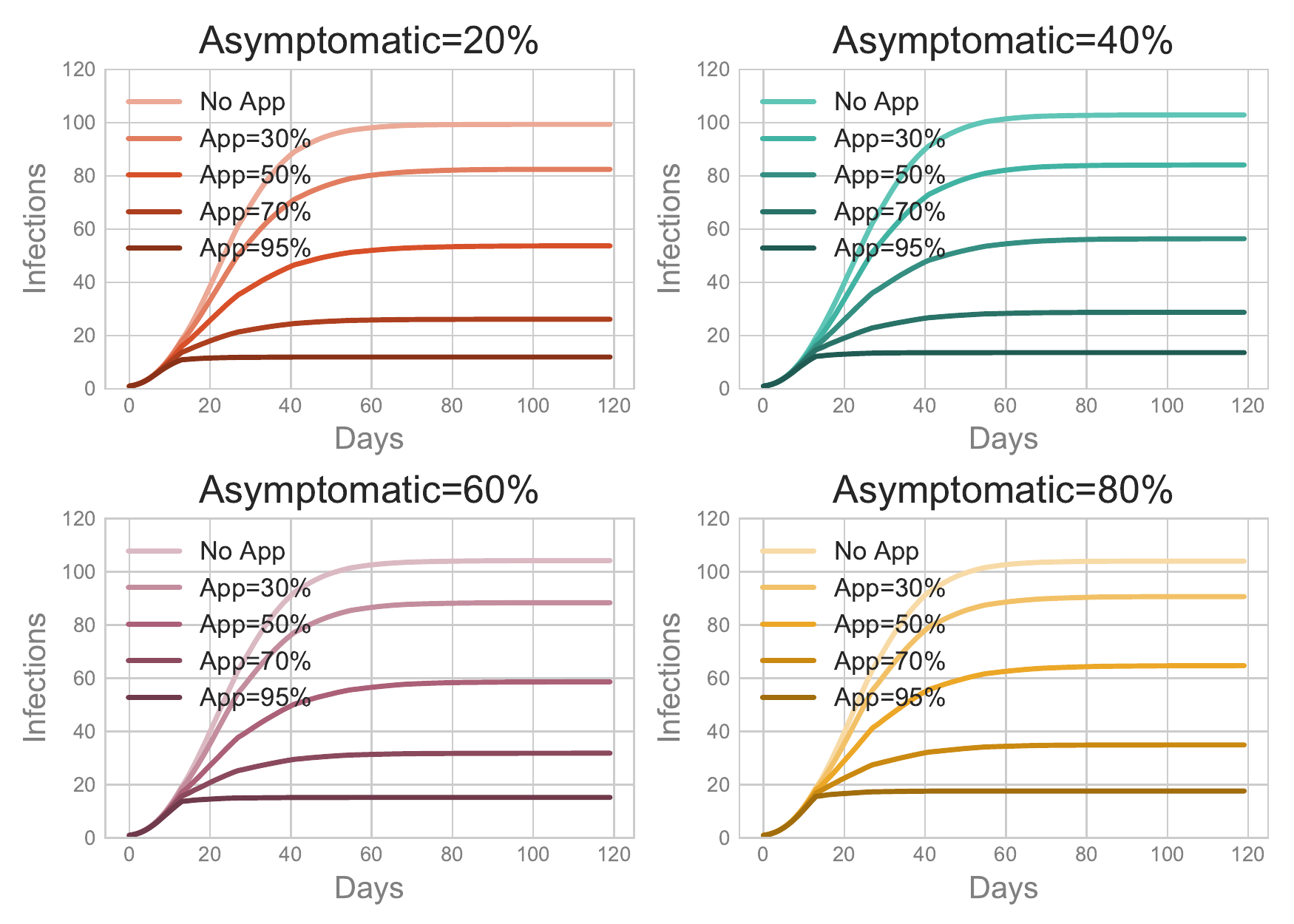}\label{S7R2Line}}
  \subfloat[Total infected but not recovered or tested.]{\includegraphics[width=8.5cm]{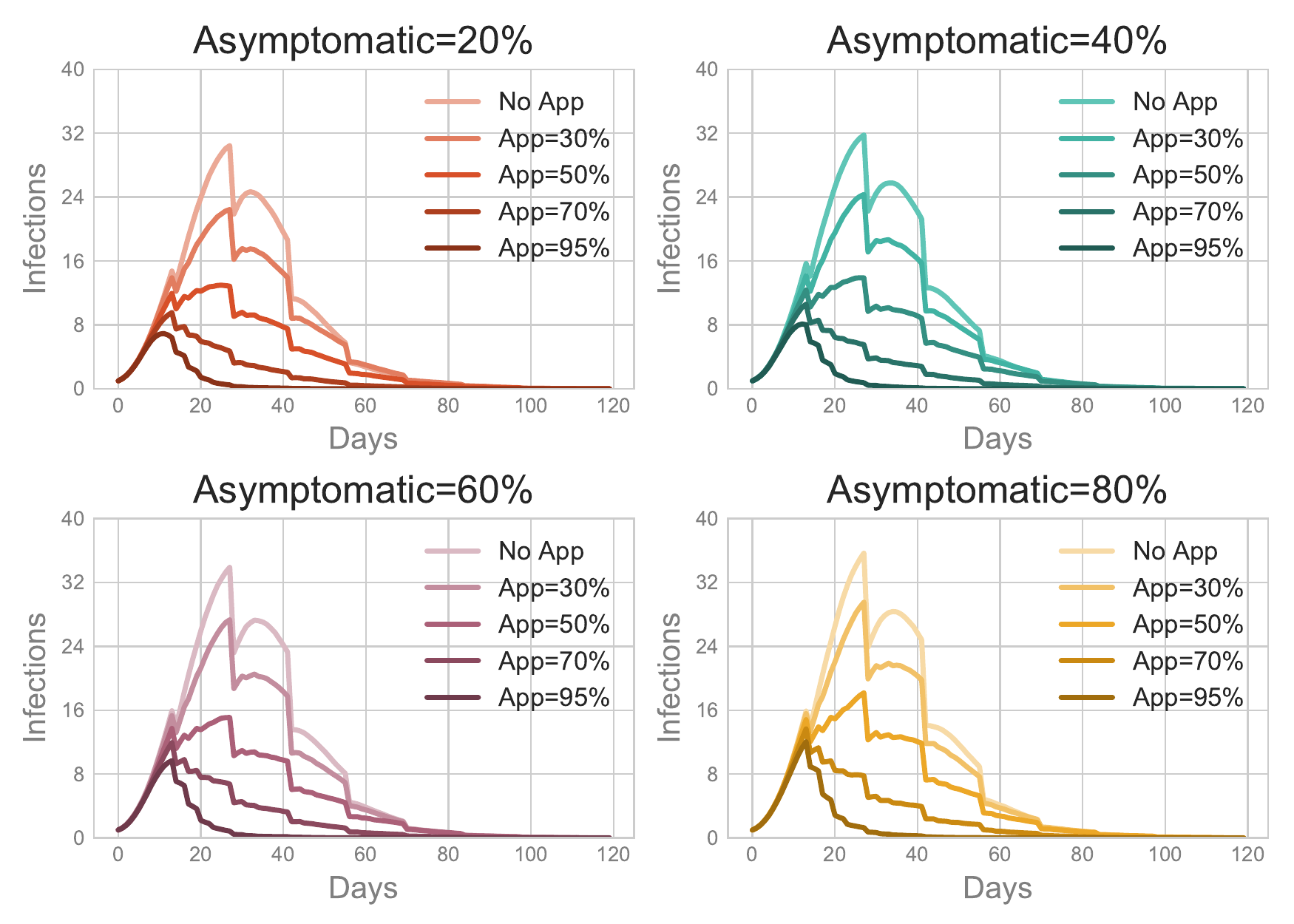}\label{S7R2Inf}}
  \caption{Pre-exposure notification system with periodic testing}
\end{figure}

In Figure \ref{S7R2Inf}, the number of infected but not tested individuals drop sharply during days when the population is tested. The graph displays bumps that come from testing the new second and third degree contacts every 3 days. Since the graphs are averaged over thousands of trials, the jaggedness is significant. It is likely caused by many second and third degree contacts being detected at the same time, and thus, the followup testing becomes synchronized. Surprisingly, periodic testing at 0\% app usage reduces infections from 106.05 in scenario 1 to 102.83, only a 3.0\% decrease. This is an insignificant change given that the population of 180 students. Thus, periodic testing without contact tracing is not effective.

\begin{table}[H]
    \caption{Pre-exposure notification system with periodic testing}
    \centering
    \begin{tabular}{|c|c|c|c|c|c|c|}
    \hline
    \multicolumn{2}{|c|}{Situation} & \multirow{ 2}{*}{Infected} & \multicolumn{3}{c|}{Quarantine} & \multirow{ 2}{*}{Tests Used}\\ 
    \cline{1-2} \cline{4-6}
    App Proportion & Asymptomatic &&False&True&Tested&\\
    \hline
    0\%&40\%&102.83 & 0 & 0 & 465.62 & 1021.34\\
    30\%&40\%&84.05 & 94.78 & 27.03 & 403.80 & 1125.57\\
    50\%&40\%&56.32 & 153.39 & 37.74 & 288.68 & 1286.85\\
    70\%&40\%&28.73 & 166.27 & 38.08 & 159.21 & 1440.73\\
    95\%&40\%&13.61 & 173.85 & 35.31 & 85.44 & 1571.49\\
    \hline
    0\%&80\%&103.95 & 0 & 0 & 375.20 & 1044.10\\
    30\%&80\%&90.64 & 104.25 & 28.62 & 358.05 & 1117.67\\
    50\%&80\%&64.69 & 174.51 & 44.14 & 281.18 & 1266.00\\
    70\%&80\%&34.95 & 195.32 & 45.23 & 172.00 & 1426.44\\
    95\%&80\%&17.66 & 208.92 & 43.91 & 105.50 & 1562.05\\
    \hline
    \end{tabular}
    \label{S7R2}
\end{table}

\subsection{Graph Extension}\label{S8}

In this section, We simulate the strategy in scenario 4 on an extended version of the graph. By the process described in Section \ref{network}, the original weighted temporal graph is extended to 5000 individuals. This population size is closer to larger communities such as schools or universities.

\begin{table}[h]
    \caption{Graph Extension}
    \centering
    \begin{tabular}{|c|c|c|c|c|c|c|}
    \hline
    \multicolumn{2}{|c|}{Situation} & \multirow{ 2}{*}{Infected} & \multicolumn{3}{c|}{Quarantine} & \multirow{ 2}{*}{Tests Used}\\ 
    \cline{1-2} \cline{4-6}
    App Proportion & Asymptomatic &&False&True&Tested&\\
    \hline
    0\%&40\%&2429.94 & 0 & 0 & 10991.94 & 31001.67\\
    30\%&40\%&1941.70 & 5858.34 & 869.33 & 10054.09 & 36557.43\\
    50\%&40\%&1753.64 & 15410.51 & 1686.31 & 9847.13 & 45625.01\\
    70\%&40\%&1301.11 & 26188.07 & 2111.04 & 7902.39 & 62003.10 \\
    95\%&40\%&501.21 & 22809.78 & 1438.89 & 3319.28 & 81969.96\\
    \hline
    0\%&80\%&2433.38 & 0 & 0 & 8759.07 & 31576.44\\
    30\%&80\%&2253.48 & 6640.62 & 1017.38 & 10021.54 & 36332.13\\
    50\%&80\%&1888.32 & 16593.59 & 1851.73 & 9563.09 & 45640.46\\
    70\%&80\%&1341.07 & 26338.16 & 2191.90 & 7671.17 & 61131.04\\
    95\%&80\%&591.61 & 26708.16 & 1723.15 & 3861.01 & 86490.97\\
    \hline
    \end{tabular}
    \label{S8R2}
\end{table}

At 40\% asymptomatic ratio, the contact tracing app reduces infections by 27.8\% at 50\% app usage and 46.5\% at 70\% app usage. This is less effective than scenario 4. At 95\% app usage, infections are reduced by 79.4\% which is similar to in scenario 4. Since the average degree is preserved on the network of 5000 students, the similar results are not surprising. As in scenario 4, the results are similar for all asymptomatic ratios. At 80\% asymptomatic ratio, the contact tracing app reduces infections by 22.4\% at 50\% app usage and 44.9\% at 70\% app usage. At 95\% app usage, infections are reduced by 75.7\%. These numbers are very similar to the corresponding values of 40\% asymptomatic ratio.

\begin{figure}[H]
  \centering
  \subfloat[Total infected.]{\includegraphics[width=8.5cm]{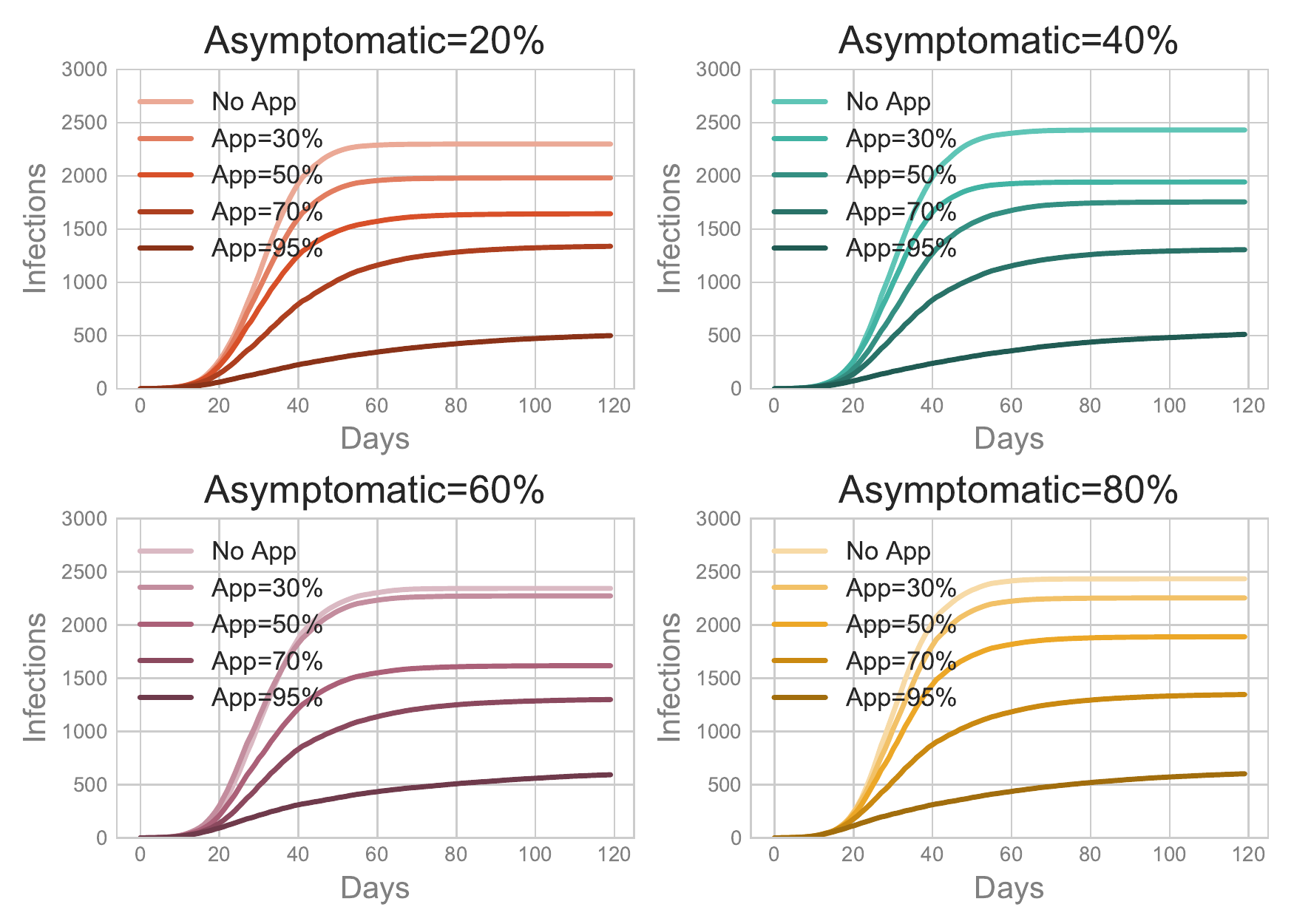}\label{S8R2Line}}
  \subfloat[Total infected but not recovered or tested.]{\includegraphics[width=8.5cm]{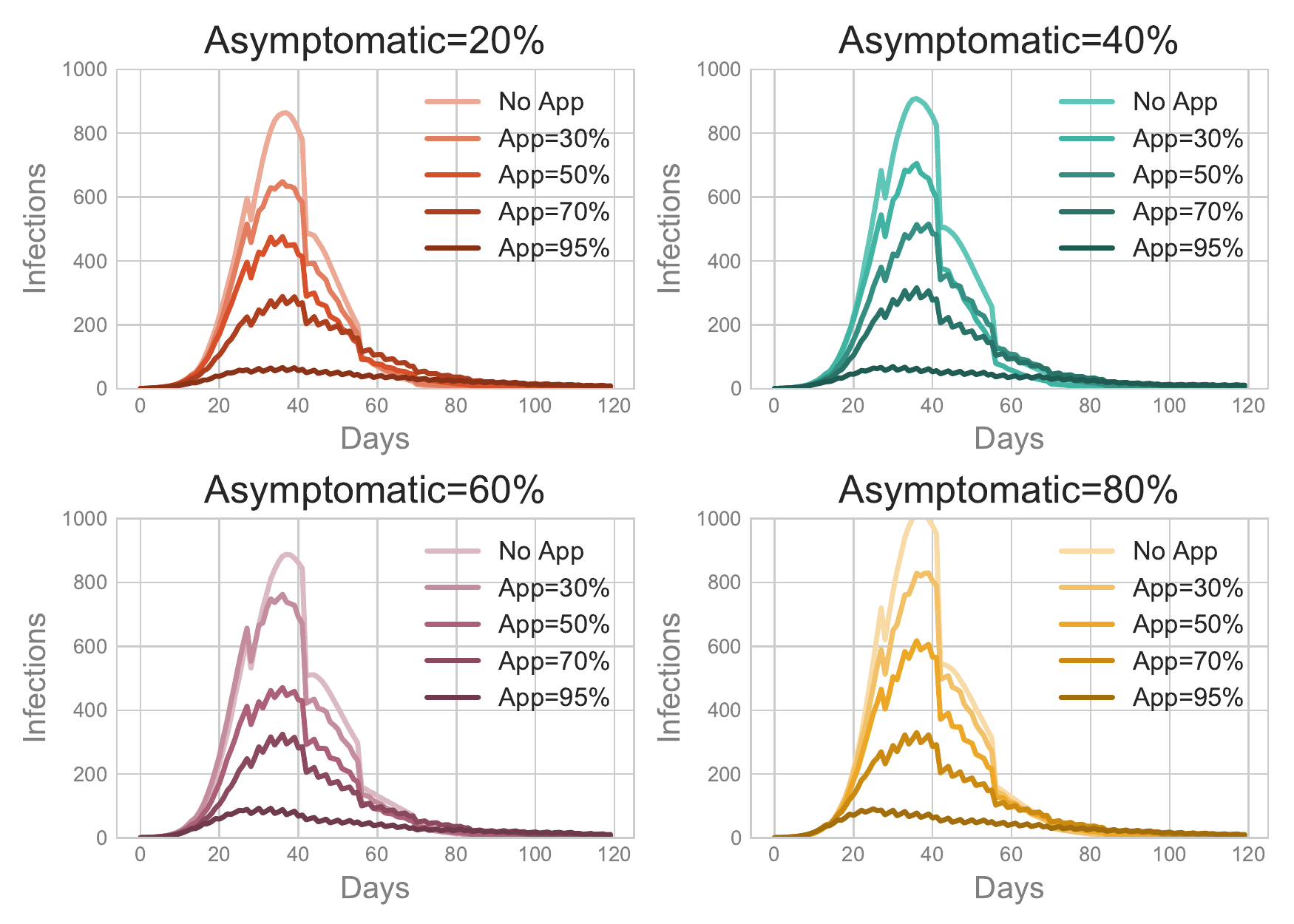}\label{S8R2Inf}}
  \caption{Graph Extension}
\end{figure}

As in scenario 4, Figure \ref{S8R2Inf} shows noticeable drops in infections on days when the population is tested. Interestingly, at 95\% app usage, the number of infected individuals never rises to more than 100 people, or 2\% of the population. The graph does not exhibit an obvious peak but rather a sustained amount of infections. Thus is a reflection of the greater population size.

\subsection{Economic Impact}\label{econ}

In this section, we estimate the economic impact of COVID-19 and preventative measures. According to a report by the Kaiser Family Foundation, most COVID-19 tests range between \$100 to \$200 with the median cost being \$127 \cite{TestCost}.
Universities are charging students up to \$1000 to quarantine for 2 weeks \cite{PurdueQ,SyracuseQ}, around \$71 a day. For students, we estimate a 20 hour work week with an hourly wage of \$20 to put daily income at \$57. The average tuition cost at universities can be as high as \$41,411 \cite{Tuition}. Assuming 180 days in a school year and that students miss classes and coursework, the average tuition per day is \$230. We roughly estimate the social cost of a single day in quarantine to be \$358 which is the sum of the previous numbers.

The economic cost of a single COVID-19 infection, or the societal willing-to-pay threshold of avoiding one COVID-19 infection, is estimated to be \$8,500 \cite{screenTesting}. We calculate the total economic cost of COVID-19 for Scenario 3 and 4. The decrease in economic cost is computed in comparison to total economic cost at 0\% app usage in scenario 1. The average economic impact per app user is the decrease in the total economic cost of COVID-19 divided by the number of app users.

\begin{table}[h]
    \caption{Average Economic Impact Per App User (Scenario 3)}
    \centering
    \begin{tabular}{|c|c|c|}
    \hline
    \multicolumn{2}{|c|}{Situation} & \multirow{2}{*}{Cost of Infection}\\ 
    \cline{1-2}
    App Proportion & Asymptomatic &\\
    \hline
    30\% & 40\% & \$991\\
    50\% & 40\% & \$2,841\\
    70\% & 40\% & \$4,260\\
    95\% & 40\% & \$4,198\\
    30\% & 80\% & -\$19\\
    50\% & 80\% & \$1,251\\
    70\% & 80\% & \$2,678\\
    95\% & 80\% & \$2,928\\
    \hline
    \end{tabular}
\end{table}

\begin{table}[h]
    \caption{Economic Value of Contact Tracing App Per User (Scenario 4)}
    \centering
    \begin{tabular}{|c|c|c|}
    \hline
    \multicolumn{2}{|c|}{Situation} & \multirow{2}{*}{Cost of Infection}\\ 
    \cline{1-2}
    App Proportion & Asymptomatic &\\
    \hline
    30\% & 40\% & -\$607\\
    50\% & 40\% & \$2,208\\
    70\% & 40\% & \$3,614\\
    95\% & 40\% & \$3,461\\
    30\% & 80\% & -\$2,052\\
    50\% & 80\% & \$974\\
    70\% & 80\% & \$2,788\\
    95\% & 80\% & \$2,927\\
    \hline
    \end{tabular}
\end{table}

The economic benefit per app user can reach as high as \$4,260 in scenario 3. In the majority of cases, the strategy in scenario 3 yields a higher benefit to user ratio. This shows that testing the entire population every 2 weeks is not cost effective. The economic impact per user is actually negative at low app usage in scenario 4. This is because the periodic testing creates significant cost while, as mentioned in Section \ref{S7}, is not effective without sufficient contact tracing.


\section{Discussion and Conclusion}\label{conclusion}

Given the significant loss of life at risk, finding effective measures to prevent the spread of COVID-19 is a top priority. Challenges in COVID-19 prevention include significant pre-symptomatic transmission, high proportions of asymptomatic cases, and inaccurate tests during the pre-symptomatic stage. This work focuses on solving these challenges by presenting effective strategies in digital contact tracing and testing.

The parameters in this model can be easily changed as more precise values of COVID-19 parameters are measured. As more data on human social patterns are collected, larger social networks can be constructed leading to more accurate predictions in the model. Additionally, as more accurate COVID-19 tests are developed, model results and optimal strategies could change. Finally, this model only considers infections based on close contacts. Although COVID-19 mainly spreads through close contacts \cite{CDCFAQ}, the significance of spread through indirect contacts is unclear and could be investigated in future models.

In conclusion, by simulating a variety of tracing and testing strategies, we found that digital contact tracing can be very effective when combined with testing. In scenario 1, the most basic strategy of quarantining first degree contacts is not effective at high asymptomatic levels, suggesting that a more comprehensive testing strategy is required. Testing second and third degree contacts greatly increases the effectiveness of the app. The pre-exposure notification system, which warns second and third degree contacts to take extra precautionary measures, can reduce infections by an additional 40\% while also reducing the number of quarantines by up to 50\%. While periodic testing with contact tracing is effective, periodic testing without contact tracing reduces infections by less than 3\%. We find the results on the extended network of 5000 students to be similar.


\section{Acknowledgements}

I would like to thank my mentor Dr. Jesse Geneson for his invaluable advice and guidance throughout the project. This would not be possible without him. Thank you to Dr. Tanya Khovanova and Alexander Vitanov for reviewing the paper draft. This work was completed under the MIT PRIMES-USA program.

\section{Appendix A: App Usage}

In the following, we present figures for the number of infections as a function of the app usage. Simulations are run with the proportion of app users ranging from 0\% to 100\% in increments of 5\%. The app becomes much more effective as more users use the app. As shown in Section \ref{S7}, scenario 4 shows very close results for all asymptomatic levels. Scenarios 3 and 4 shows that the effectiveness of the app begins to level off after more than 80\% of the population uses the app.

\begin{figure}[H]
\renewcommand\thefigure{A.1} 
  \centering
  \subfloat[Traditional App Configuration]{\includegraphics[width=8.5cm]{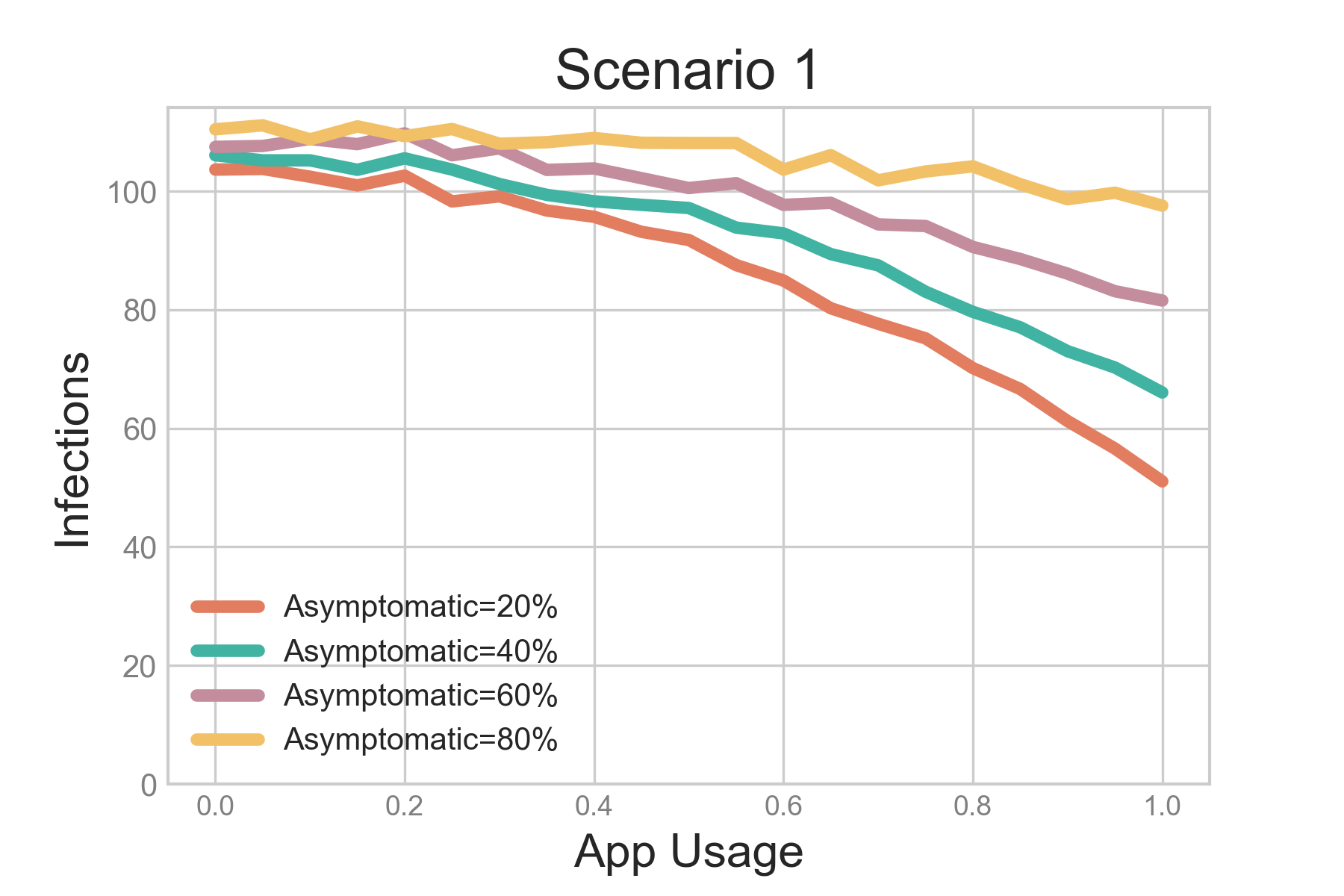}\label{S1B}}
  \subfloat[Testing second and third degree contacts]{\includegraphics[width=8.5cm]{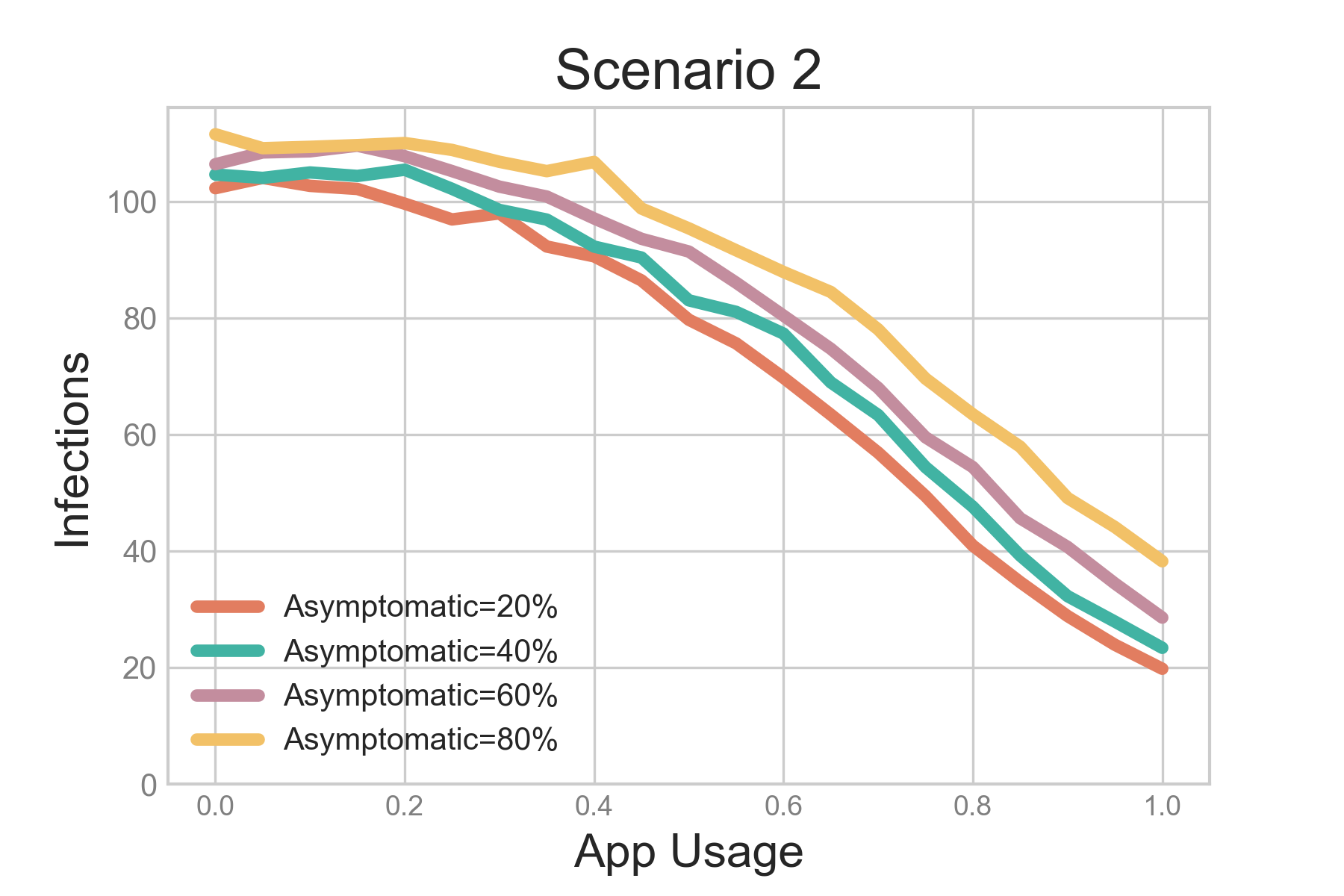}\label{S2B}}
  \newline
  \subfloat[Pre-exposure notification system]{\includegraphics[width=8.5cm]{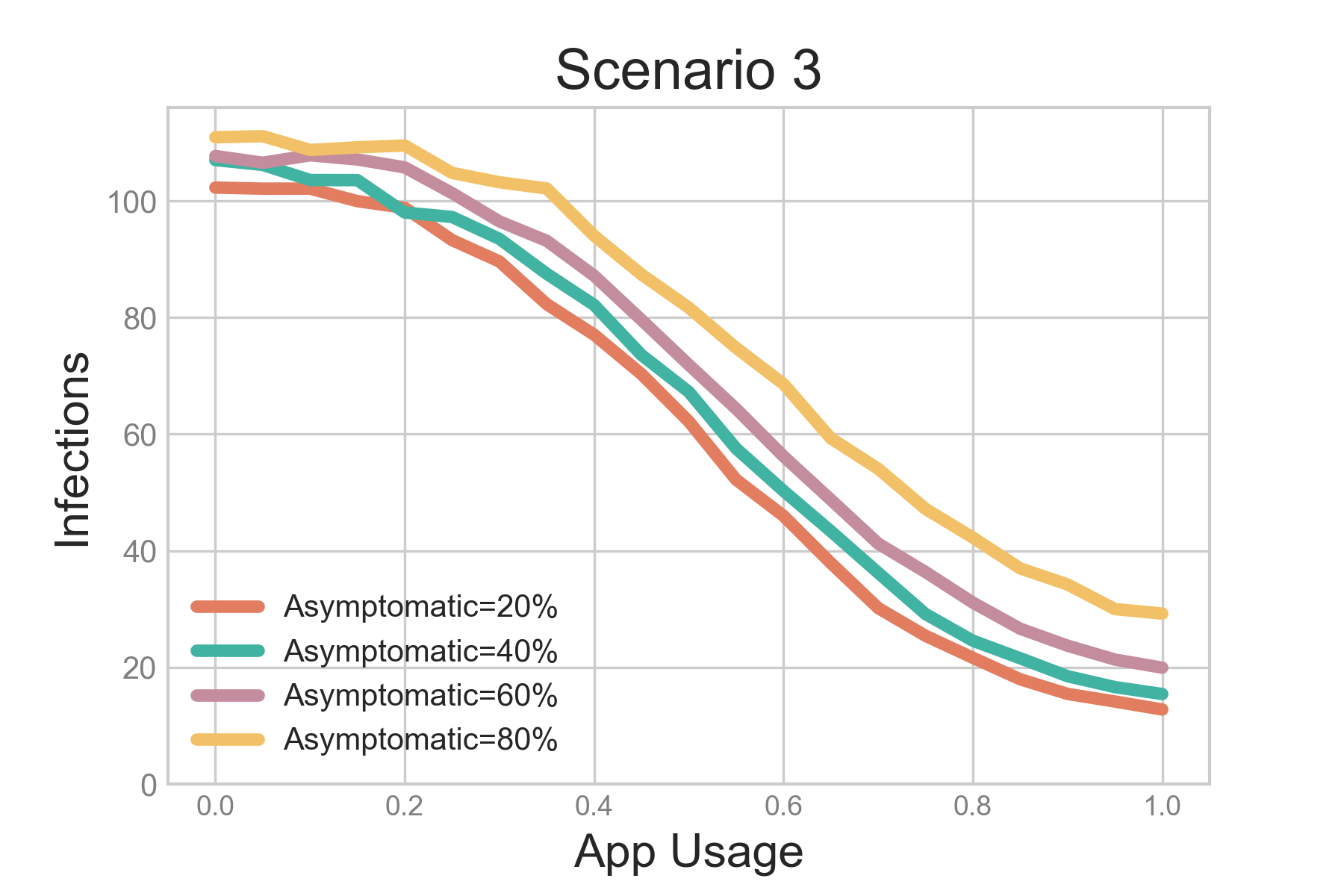}\label{S3B}}
  \subfloat[Periodic testing]{\includegraphics[width=8.5cm]{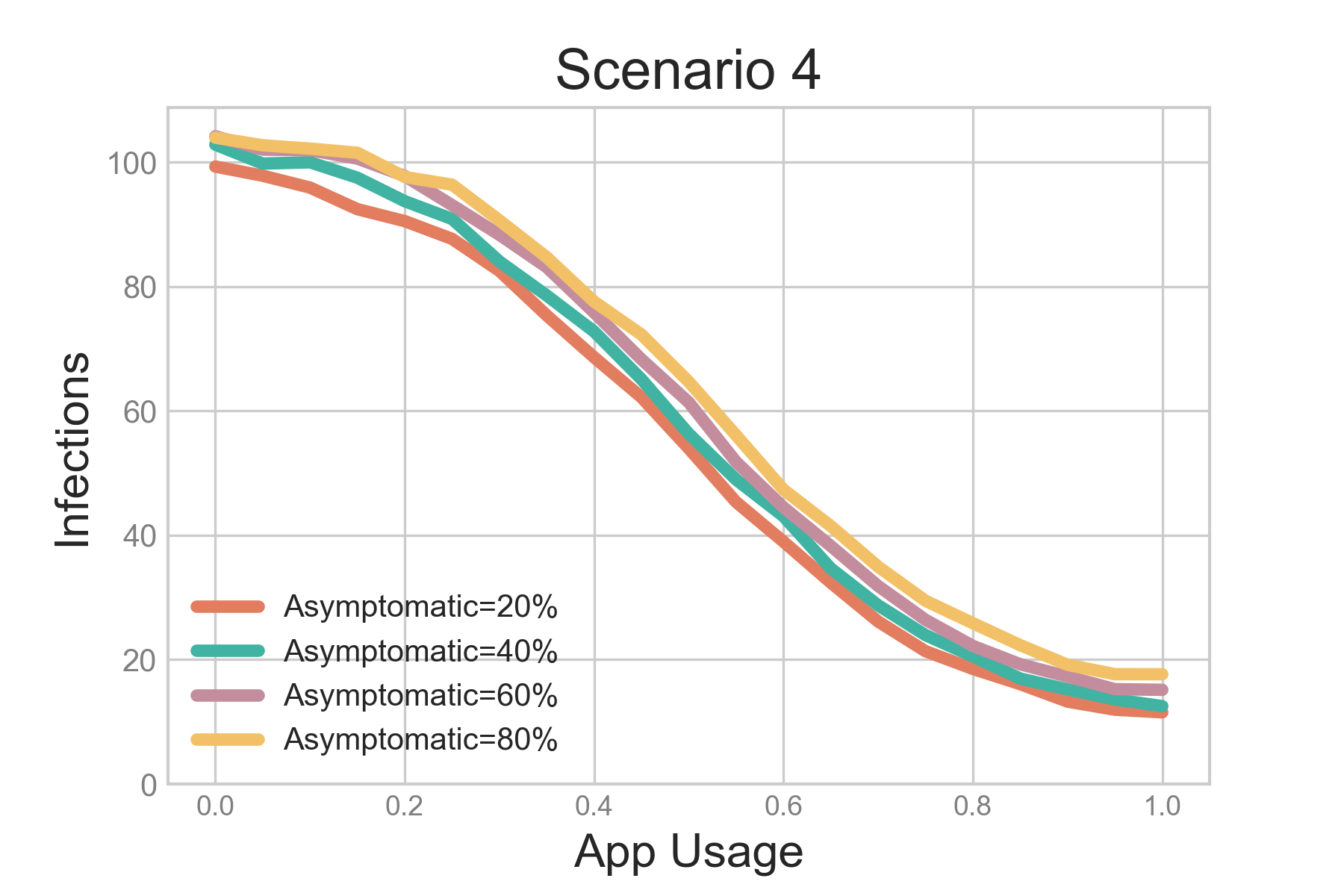}\label{S4B}}
  \caption{App Usage vs Infections}
\end{figure}


\bibliographystyle{plain}
\bibliography{mybibfile}

\end{document}